# Van der Waals metal−semiconductor junction: weak Fermi level pinning enables effective tuning of Schottky barrier

Yuanyue Liu,* Paul Stradins and Su-Huai Wei*

National Renewable Energy Laboratory, Golden, CO, USA
*Correspondence to: yuanyue.liu.microman@gmail.com, suhuaiwei@csrc.ac.cn

**Abstract:** Two-dimensional (2D) semiconductors have shown great promise in (opto)electronic applications. However, their developments are limited by a large Schottky barrier (SB) at the metal–semiconductor junction (MSJ), which is difficult to tune by using conventional metals due to the strong Fermi level pinning (FLP) effect. Here we show that, this problem can be overcome by using 2D metals, which are bounded with 2D semiconductors through van der Waals (vdW) interaction. This success relies on a weak FLP at the vdW MSJ, which is attributed to the suppression of metal-induced gap states. Consequently, the SB becomes tunable and can vanish with proper 2D metals (e.g. *H*-NbS$_2$). This work not only offers new insights into the fundamental properties of heterojunctions, but also uncovers great potential of 2D metals in device applications.

**Main Text:** Metal–semiconductor junction (MSJ) is a critical component in (opto)electronic devices. One of the most important parameters for the MSJ is the Schottky barrier (SB) height ($\Phi$), an energy barrier height for charge carriers transport across the junction, which has significant impact on device performance.[1-3] The importance of tuning $\Phi$ has recently been emphasized in the devices based on two-dimensional (2D) semiconducting transition metal dichalcogenides (MX$_2$ where M = Mo or W, and X = S, Se, or Te), which have attracted intense interest due to their unique properties and promising applications.[4-7] Tuning $\Phi$ could enable a variety of improvements, such as reducing the contact resistance, modulating the carrier polarity in the channel for transistors, and enhancing the selectivity of carrier extraction for photovoltaic cells. The first one is of particular importance to match the performance with Si or III-V device analogues.[8, 9]

However, it is rather difficult to tune $\Phi$ for 2D MX$_2$ by using different common metals, due to the effect of Fermi level pinning (FLP). The reason is the following: generally, $\Phi$ is determined by the energy difference between the Fermi level (FL) and semiconductor band edges in the junction:

$\Phi_e = E_{CBM} - E_F$ and $\Phi_h = E_F - E_{VBM}$ (1)

where $\Phi_e$ and $\Phi_h$ are the barrier heights for electrons and holes, respectively; $E_F$ is the Fermi energy; and $E_{VBM}$ and $E_{CBM}$ are the energy of the valence band maximum (VBM) and conduction band minimum (CBM), respectively, of the semiconductor in the junction. Ideally, neglecting the metal−semiconductor interaction, the $\Phi$ should follow the predictions of the Schottky-Mott model:

$$\Phi_e = W - E_{ea} \text{ and } \Phi_h = E_{ip} - W \qquad (2)$$

where $W$ is the work function of the metal, $E_{ea}$, and $E_{ip}$ are the electron affinity and ionization potential of the semiconductor, respectively. $E_{ip} - E_{ea}$ = band gap. These quantities are the intrinsic properties of isolated materials before forming the junction. The Schottky–Mott model suggests that the $\Phi_e$ (or $\Phi_h$) is linearly dependent on the $W$ of metals with a slope of +/−1. However, in reality, the $\Phi$ is usually insensitive to $W$, and the FL of the system is pinned to a fixed position in the semiconductor band gap, varying little with respect to different metals used. The strength of FLP for a given semiconductor with a set of metals can be characterized by the linearly-fitted slope of the $\Phi$ vs $W$ plot:[2]

$$S = |\,d\Phi\,/\,dW\,| \qquad (3)$$

If $S = 1$, the Schottky–Mott limit is recovered. Unfortunately, S is small for 3D metal−2D semiconductor junctions. Experiments measure the $S = 0.1$ for 3D metal−$MoS_2$ junction,[10] indicating a strong FLP.

In general, there exist many different models for explaining the FLP (see Ref.[2] for a review), and no simple equations are applicable for all type of MSJs.[2] Here we focus on those which have been recognized to be relevant for 2D semiconductors: (i) Formation of metal-induced gap states (MIGS) in the semiconductor[11-14], as also observed in many 3D metal−3D semiconductor junctions.[2, 15-20] These states serve as reservoir for electrons or holes and therefore pin the FL. (ii) Interface dipole formed by the charge redistribution at the interface can shift the electronic levels from their original positions, leading to a deviation from the Schottky-Mott limit[2, 11, 21, 22]. (iii) Defects[23] at the interface (created during materials/device fabrication) could generate gap states that pin the FL,[1, 2] yet this can be neglected for a high-quality interface.

In this work, we show that, contrary to the conventional 3D metal−2D semiconductor junction, the FLP is weak for the MSJ formed by van der Waals (vdW) interactions, which is attributed to the suppression of MIGS in the semiconductor. This phenomenon allows for the tuning of $\Phi$ by using different 2D metals. Based on this, we identify the promising 2D metals that can form a low-Φ junction with the 2D semiconductors.

We consider a wide range of 2D metals that have been experimentally realized, including: (i) triangular ($T$) and distorted triangular ($T'$) phases of $MoX_2$ and $WX_2$; (ii) group-5 $MX_2$ (M = V, Nb, Ta) and $TiX_2$, which exhibit either hexagonal ($H$) or $T$ phase at room temperature; and (iii) pristine and doped graphene. These materials have chemically saturated surfaces and bind with the 2D semiconductors through vdW interactions.

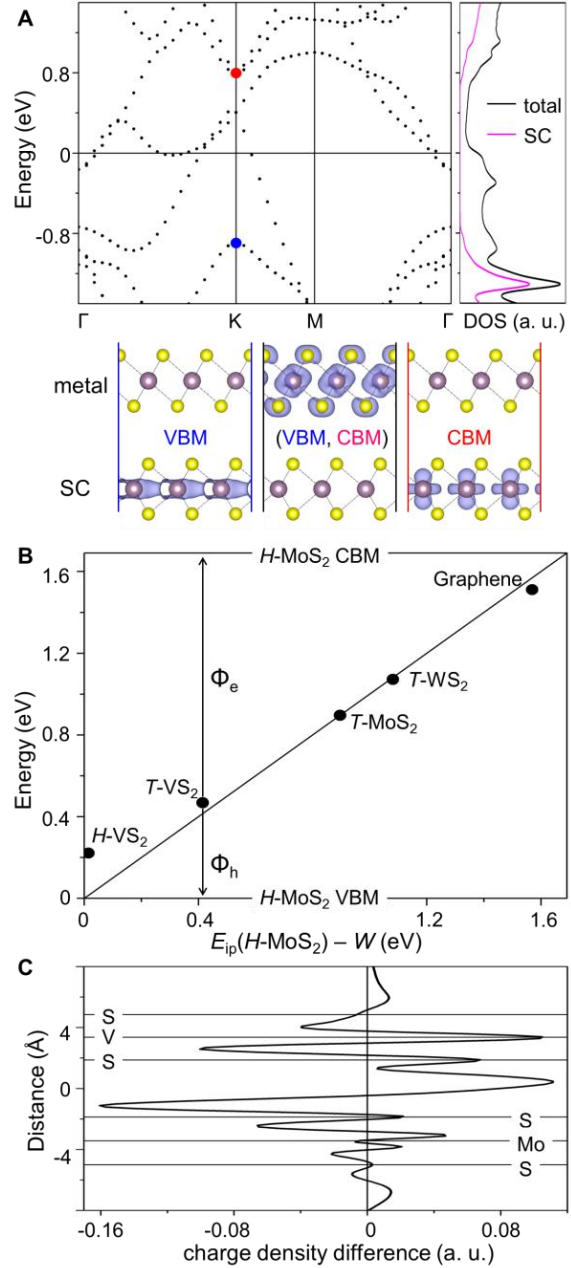

**Figure 1**. Weak FLP at the vdW MSJ. (A): Electronic band structure and density of states (DOS) of a typical vdW MSJ. $T$-MoS$_2$−$H$-MoS$_2$ is used here as an example, where $T$-MoS$_2$ is a metal and $H$-MoS$_2$ is a semiconductor (SC). Fermi level is set to zero. The VBM and CBM of the SC are marked by the blue and red dots, respectively. Purple line shows the DOS projected on the semiconductor. Isosurfaces show the spatial distributions of the states. (B): SB heights ($\Phi_e$ for electrons and $\Phi_h$ for holes) between $H$-MoS$_2$ and various 2D metals. The diagonal line shows the values predicted by the Schottky-Mott model. (C) Charge density change (averaged in the plane parallel to the interface) after forming the MSJ.

The atom positions perpendicular to the basal plane are shown as the distances relative to the center of the interface, and marked by the horizontal lines.

Fig. 1a shows a typical band structure of a 2D MSJ by using $T$-MoS$_2$−$H$-MoS$_2$ as an example ($T$-MoS$_2$ is a metal and $H$-MoS$_2$ is a semiconductor; see computational details in the Supporting Information). The origin of each electronic state can be determined from its charge density distribution, based on which we identify the VBM and CBM of the semiconductor in the junction, as marked by the blue and red dots respectively in Fig. 1a. We then plot the distribution of the states between the VBM and the CBM in the junction (Fig. 1a), and find that these states come from the metal rather than the semiconductor. Therefore, the MIGS in the semiconductor of the junction are negligible, which is further evidenced by projecting density of states (DOS) onto the semiconductor (Fig. 1a). This feature is in contrast to the 3D metal−2D semiconductor junctions, where MIGS are significant[11-14]. We attribute this phenomenon to the unique interaction between the metal and the semiconductor at the vdW MSJ, which is substantially weaker than the chemical bonding in other junctions.[11, 14] It is interesting to notice that the interaction can also be weakened by inserting a thin layer between the metal and the semiconductor to break their direct chemical bonding.[24, 25] However, in these cases, a proper separation material has to be chosen first. This need is avoided in our case, which also simplifies the device fabrication.

The suppression of MIGS in the semiconductor can therefore lead to the weak FLP. To verify it, Fig. 1b shows the $\Phi$ (calculated by using Eq. 1) for $H$-MoS$_2$ with various 2D metals. We choose $T$-MoS$_2$, $T$-WS$_2$, $H$- and $T$-VS$_2$ to study because of their small lattice mismatch (<2%) with $H$-MoS$_2$, making the computations efficient. Graphene (though strictly speaking, a semi-metal) is also selected because it has been used experimentally as an electrode for $H$-MoS$_2$[26-30] (though a large supercell with commensurate structure is required for modeling; see Fig. S1 in the SI for the structure model; the commensurate structures are also used for other largely lattice-mismatched junctions as shown later). Notably, we find that the $\Phi$ largely follows the trend of the Schottky-Mott limit (the diagonal line) and its value can vary in a wide range, allowing the tuning of $\Phi$ by using different 2D metals. Although different computational method (more specifically, density-functional) yields different $S$, they all give a higher $S$ compared with that of 3D metal−$H$-MoS$_2$ junctions (see Fig. S2, S3 and the related text in the SI), indicating a weaker FLP.

It has to be noted that, although the vdW interaction is weak, it can redistribute the charge density at the interface and give rise to an interface dipole at the MSJ. This is shown in Fig. 1c, where the charge density ($\rho$) change after forming the junction is calculated for the example of $H$-$VS_2$−$H$-$MoS_2$:

$$\Delta\rho = \rho(\text{junction}) - \rho(\text{metal}) - \rho(\text{semiconductor}) \quad (4)$$

We find an asymmetric charge accumulation/depletion across the interface, indicating the formation of an interface dipole. As discussed above, this interface dipole can shift the electronic levels from their original positions, leading to a deviation from the Schottky-Mott limit.[2, 11, 21, 22]

Since the FLP is weak for the vdW junction, the $\Phi_h$ (or $\Phi_e$) can be effectively reduced by using high-$W$ (or low-$W$) metals as electrodes. To identify the promising candidates, we calculate the $W$ for various 2D metals, and compare them with the $E_{ip}$ and $E_{ea}$ of 2D semiconductors, as shown in Figure 2. Interestingly, these energies show a systematic variation as the cation or anion changes.

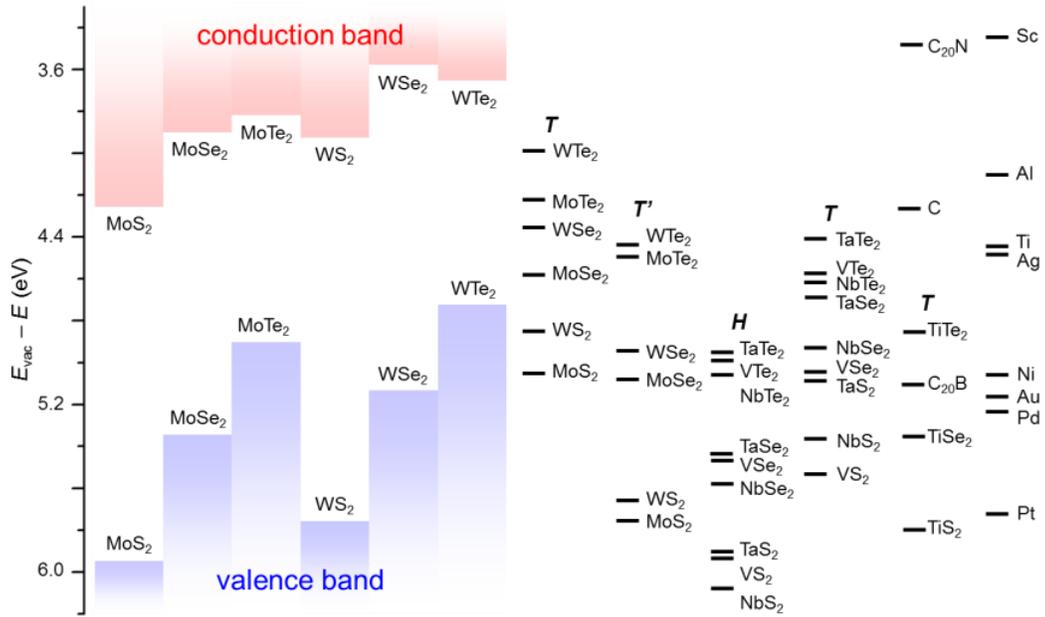

**Figure 2**. Band alignment between 2D metals and semiconductors. Left columns show the electron affinity and ionization potential of semiconductors. Right bars show the work function of metals. The phase is labeled in italic. C stands for pristine graphene, C20N is the N-doped graphene with the C:N = 20:1, and similarly for C20B. For comparison, the work functions of some commonly used 3D metals are also shown.

For 2D semiconductors, we find that (i) as the atomic number of X increases (from S to Se to Te), the CBM and VBM energies increase (except that the CBM of WTe$_2$ is lower than that of WSe$_2$), and band gap decreases. (ii) Moreover, for the common-X system, the CBM and VBM energies of WX$_2$ are higher than those of MoX$_2$. These observations agree with literature results calculated using different methods.[31, 32] Therefore, MoS$_2$ has the lowest CBM and VBM, WTe$_2$ has the highest VBM, and WSe$_2$ has the highest CBM. These results explain why MoS$_2$ is usually n-type in experiments but WSe$_2$ is p-type: compared with MoS$_2$, WSe$_2$ has a higher CBM and VBM and therefore has a higher $\Phi_e$ but lower $\Phi_h$ for given metals, resulting in an easier hole injection but more difficult electron injection.[33]

For 2D metals, we find that (i) for the same M, the $W$ of MX$_2$ increases as the X atomic number decreases; and (ii) for the same X, the $W$ increases as the M atomic number (in the same group) decreases (except VX$_2$ that is magnetic, in which case the spin-polarization changed the order between VX$_2$ and NbX$_2$). These trends originate from the coupling between the M $d$ and X $p$ orbitals. As shown by the projected DOS (Figures 3a), the states near the FL are contributed by the M $d$ and X $p$ states, and their spatial distribution suggests they are antibonding-coupled (Figure 3b). For common-M systems, because M $d$ states have higher energy than X $p$ states[31] (Figure 3c), as the $p$ energy decreases from Te to Se to S, the $p$-$d$ coupling becomes weakened and the level repulsion reduces. This decreases the energy of the antibonding states, and thus, the FL moves down and $W$ increases. For common-X systems, the M $d$ energy decreases and the Coulomb repulsion between the $d$ electrons increases when M moves up in the same group;[34] thus, the FL decreases and $W$ increases. We also find that for different phases of MX$_2$, the FL decreases as the stability increases. For example, $T'$-MoX$_2$ and WX$_2$ are more stable than the corresponding $T$ phases[35] and have lower FLs; the situation is similar for group-5 $H$-MX$_2$ compared with their $T$ phases. Notably, we find that some of the metallic MX$_2$ (group-5 $H$-MX$_2$ and TiS$_2$) have higher $W$ than Pt, which possesses the highest $W$ among elemental metals. Of these, $H$-NbS$_2$ has the highest $W$, suggesting it could be a promising electrode to achieve low-$\Phi_h$ contact with the semiconductor, as shown later.

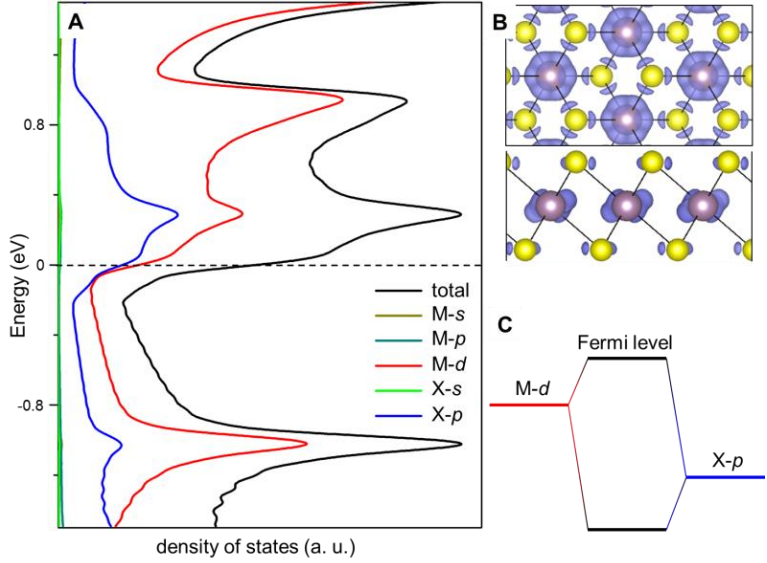

**Figure 3**. (**A**): Density of states (DOS) of metallic $MX_2$ ($T$-$MoS_2$ is used here as an example). Fermi level is set to zero. The black line shows the total DOS, and the others are the projected DOS on the orbitals of M and X. (**B**) Charge density distribution of the states in the (-0.025, 0.025) eV range (top and side views). (**C**) Schematic of coupling between M $d$ states and X $p$ states.

We also explore the potential of using doped graphene as electrodes. B has fewer electrons than C, thus B doping increases the $W$ of graphene; Inversely, N has more electrons than C and its doping decreases the $W$. In particular, at high N doping concentration (C:N = 20:1), the $W$ is as low as that of Sc and below all 2D $MX_2$ metals, implying that it could be a promising electrode for electron injection.

After identifying the promising metal electrodes based on their $W$, we calculate their $\Phi$ with the 2D semiconductors, as shown in Fig. 4 (see Fig. S4 for representative band structures, and Table S1 for comparison of the results calculated by using different methods). Indeed, $H$-$NbS_2$ can form a low-$\Phi_h$ junction. In particular, the $\Phi_h$ is negative when $H$-$NbS_2$ contacts $H$-$WTe_2$, $MoTe_2$, $WSe_2$ or $MoSe_2$, indicating a spontaneous electron transfer to $H$-$NbS_2$ and hole injection to the semiconductor upon contact. Note that this strong charge transfer shifts the FL and the VBM from their original positions, leading to a deviation from the ideal Schottky-Mott model. Similarly, $C_{20}N$ can form a low-$\Phi_e$ contact. However, different from $H$-$NbS_2$, which has been experimentally synthesized, an efficient N-doping approach up to a high concentration has to be developed.

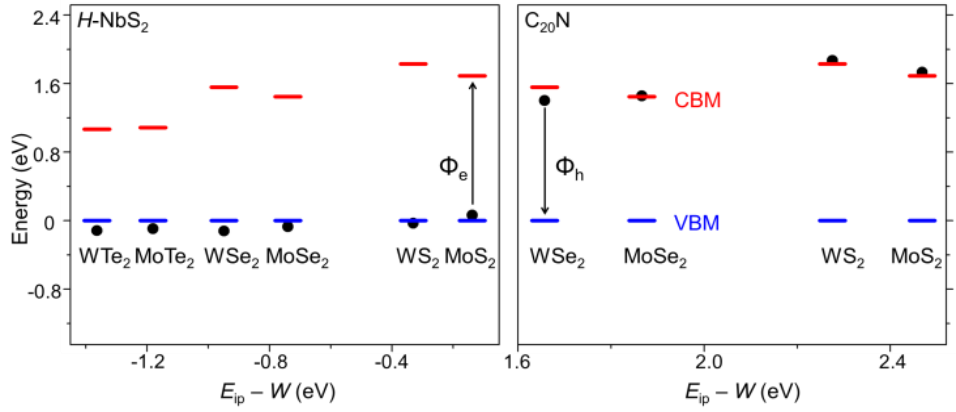

**Figure 4**. SB heights for $H$-NbS$_2$ and C$_{20}$N metals, with various 2D semiconductors. C$_{20}$N−tellurides are not calculated because of the extremely large supercell required for modeling to reduce the lattice mismatch. The red bars indicate the CBM, and the blue bars represent the VBM that are set to zero.

There are other benefits of using 2D metal as an electrode. Its transparency and flexibility are compatible with similar features of the 2D semiconductor channel, and it can be integrated for transparent and flexible electronics.[28, 30] Second, 2D metal has limited electronic density of states and therefore low quantum capacitance. Thus, when charge is accumulated by applying a dielectric-mediated voltage, its $W$ changes dramatically compared with the conventional metals, which have high density of states. This unique feature of 2D metal leads to a gate-tunable $W$ and therefore Φ, as has been observed in experiments.[26, 36-39] Third, it has been shown that a flat interface between the metal and semiconductor could help carrier transport.[40] The atomically flat interface is difficult to achieve by using conventional metals, but can be easily realized at the vdW contact. In addition, Using 2D metal as electrodes allows full encapsulation of the semiconductor, which prevents contamination from the environment,[29, 39] or stabilize the reactive semiconductor such as phosphorene.[41] Moreover, the suppression of MIGS reduces the electron-hole recombination at the interface, leading to a higher energy conversion efficiency for optoelectronic devices.

We also point out that although the focus of this work is on the vdW junction, the band alignment shown in Figure 2 could also provide guidelines for designing a low-resistance "edge" contact junction, where the 2D metals and semiconductors are chemically bonded in the same basal plane.[42] In addition, Figure 2 also offers guidelines for designing photovoltaic cells based on 2D materials; in this case, it is critical to select electrodes with a FL close to the VBM or CBM of the semiconductor to obtain a high open-circuit voltage.

In summary, the problem of large and untunable SB is overcome by replacing the conventional metals with 2D metals, which form vdW MSJ with 2D semiconductors. This type of MSJ exhibits weak FLP, which is attributed to the suppression of MIGS. This work not only advances the understanding of fundamental properties of heterojunctions, but also shows novel functions of 2D metals in device applications.

**Acknowledgement**

Y. L. acknowledges discussions with Boris Yakobson and Brandon Wood on $NbS_2$. This research was funded by the U.S. Department of Energy under Contract No. DE-AC36-08GO28308 with the National Renewable Energy Laboratory (NREL). This work used computational resources at NREL and NERSC (supported by DOE Office of Science DE-AC02-05CH11231).